\documentstyle[11pt,psfig,twoside]{article}

\begin{document}

\newcommand{\dd}{\,{\rm d}}
\newcommand{\ie}{{\it i.e.},\,}
\newcommand{\etal}{{\it et al.\ }}
\newcommand{\eg}{{\it e.g.},\,}
\newcommand{\cf}{{\it cf.\ }}
\newcommand{\vs}{{\it vs.\ }}
\newcommand{\zdot}{\makebox[0pt][l]{.}}
\newcommand{\up}[1]{\ifmmode^{\rm #1}\else$^{\rm #1}$\fi}
\newcommand{\dn}[1]{\ifmmode_{\rm #1}\else$_{\rm #1}$\fi}
\newcommand{\upd}{\up{d}}
\newcommand{\uph}{\up{h}}
\newcommand{\upm}{\up{m}}
\newcommand{\ups}{\up{s}}
\newcommand{\arcd}{\ifmmode^{\circ}\else$^{\circ}$\fi}
\newcommand{\arcm}{\ifmmode{'}\else$'$\fi}
\newcommand{\arcs}{\ifmmode{''}\else$''$\fi}
\newcommand{\MS}{{\rm M}\ifmmode_{\odot}\else$_{\odot}$\fi}
\newcommand{\RS}{{\rm R}\ifmmode_{\odot}\else$_{\odot}$\fi}
\newcommand{\LS}{{\rm L}\ifmmode_{\odot}\else$_{\odot}$\fi}

\newcommand{\Abstract}[2]{{\footnotesize\begin{center}ABSTRACT\end{center}
\vspace{1mm}\par#1\par
\noindent
{\bf Key words:~~}{\it #2}}}

\newcommand{\TabCap}[2]{\begin{center}\parbox[t]{#1}{\begin{center}
  \small {\spaceskip 2pt plus 1pt minus 1pt T a b l e}
  \refstepcounter{table}\thetable \\[2mm]
  \footnotesize #2 \end{center}}\end{center}}

\newcommand{\TableSep}[2]{\begin{table}[p]\vspace{#1}
\TabCap{#2}\end{table}}

\newcommand{\TableFont}{\footnotesize}
\newcommand{\TableFontIt}{\ttit}
\newcommand{\SetTableFont}[1]{\renewcommand{\TableFont}{#1}}

\newcommand{\MakeTable}[4]{\begin{table}[htb]\TabCap{#2}{#3}
  \begin{center} \TableFont \begin{tabular}{#1} #4 
  \end{tabular}\end{center}\end{table}}

\newcommand{\MakeTableSep}[4]{\begin{table}[p]\TabCap{#2}{#3}
  \begin{center} \TableFont \begin{tabular}{#1} #4 
  \end{tabular}\end{center}\end{table}}

\newenvironment{references}%
{
\footnotesize \frenchspacing
\renewcommand{\thesection}{}
\renewcommand{\in}{{\rm in }}
\renewcommand{\AA}{Astron.\ Astrophys.}
\newcommand{\AAS}{Astron.~Astrophys.~Suppl.~Ser.}
\newcommand{\ApJ}{Astrophys.\ J.}
\newcommand{\ApJS}{Astrophys.\ J.~Suppl.~Ser.}
\newcommand{\ApJL}{Astrophys.\ J.~Letters}
\newcommand{\AJ}{Astron.\ J.}
\newcommand{\IBVS}{IBVS}
\newcommand{\PASP}{P.A.S.P.}
\newcommand{\Acta}{Acta Astron.}
\newcommand{\MNRAS}{MNRAS}
\renewcommand{\and}{{\rm and }}
\section{{\rm REFERENCES}}
\sloppy \hyphenpenalty10000
\begin{list}{}{\leftmargin1cm\listparindent-1cm
\itemindent\listparindent\parsep0pt\itemsep0pt}}%
{\end{list}\vspace{2mm}}

\def\TYLDA{~}
\newlength{\DW}
\settowidth{\DW}{0}
\newcommand{\dw}{\hspace{\DW}}

\newcommand{\refitem}[5]{\item[]{#1} #2%
\def\REFARG{#3}\ifx\REFARG\TYLDA\else, {\it#3}\fi
\def\REFARG{#4}\ifx\REFARG\TYLDA\else, {\bf#4}\fi
\def\REFARG{#5}\ifx\REFARG\TYLDA\else, {#5}\fi.}

\newcommand{\Section}[1]{\section{#1}}
\newcommand{\Subsection}[1]{\subsection{#1}}
\newcommand{\Acknow}[1]{\par\vspace{5mm}{\bf Acknowledgements.} #1}
\pagestyle{myheadings}

\def\thefootnote{\fnsymbol{footnote}}

\begin{center}
{\Large\bf The Optical Gravitational Lensing Experiment.\\
\vskip3pt
Population Effects on the Mean Brightness\\
\vskip3pt
of the Red Clump Stars
\footnote{Based on  observations obtained with the 1.3~m Warsaw
telescope at the Las Campanas  Observatory of the Carnegie Institution
of Washington.}}
\vskip1cm
{\bf A.~~U~d~a~l~s~k~i}
\vskip5mm
{Warsaw University Observatory, Al.~Ujazdowskie~4, 00-478~Warszawa,
Poland\\
e-mail: udalski@sirius.astrouw.edu.pl }
\end{center}

\Abstract{We present an empirical test indicating that the mean {\it
I}-band magnitude of the red clump stars, used as the standard candle in
the recent distance determinations to the Magellanic Clouds and other
objects, is age independent for intermediate age ($2-10$~Gyr) stars.
Fifteen star clusters of age $\approx 1.5-12$~Gyr from the LMC and SMC
(ESO121SC03, SL663, NGC~2155, NGC~2121, SL388, SL862, NGC~121, L1,
KRON3, NGC~416, L113, NGC~339, L11, NGC~419, NGC~411) were observed and
their color-magnitude diagrams are presented. The mean {\it I}-band
brightness of the red clump in these clusters is constant and its mean
extinction-free magnitude is: $I_0=17.88\pm0.05$ mag and $I_0=18.31\pm
0.07$ mag at the mean metallicity of $-0.8$~dex and $-1.2$~dex for the
LMC and SMC clusters, respectively. For older objects ($> 10$~Gyr) the
brightness of the red clump, which converts into the red part of the
horizontal branch, fades by about $0.3-0.4$~mag, setting an important
limitation on the red clump stars method of distance determination.

The red clump distance moduli to the Magellanic Clouds from the new
independent data set are: $m-M=18.18\pm0.06$ mag and $m-M=18.65\pm0.08$
mag for the LMC and SMC, respectively, in very good agreement with
previous determinations.

Weak dependence of the mean {\it I}-band brightness of the red clump on
metallicity and its independence of age for intermediate age population
($2-10$~Gyr) of stars as well as the most precise calibration as compared
to other standard candle candidates makes the red clump stars method one
of the most accurate steps in the distance scale ladder.}{Magellanic
Clouds -- Galaxies: distances and redshifts -- distance scale --
clusters individual: ESO121SC03, SL663, NGC~2155, NGC~2121, SL388,
SL862, NGC~121, L1, KRON3, NGC~416, L113, NGC~339, L11, NGC~419, NGC~411} 

\Section{Introduction}

In a series of papers (Udalski \etal 1998a, Udalski 1998) a new
determination of distance to the Magellanic Clouds was presented. It was
based on the newly developed red clump stars method proposed by
Paczy{\'n}ski and Stanek (1998). The resulting distance modulus to the
LMC turned out to be $m-M\approx18.1$ mag giving the distance about 15\%
smaller than generally accepted value from Cepheid variables. It was,
however, in excellent agreement with the distance determination from
RR~Lyr stars (Udalski 1998), providing a strong argument in favor of the
"short" distance scale to the LMC. Results of Udalski \etal (1998a)
towards the LMC were confirmed  by Stanek, Zaritsky and Harris (1998)
based on independent photometry and extinction determination. Also the
distance to the SMC turned out to be shorter than previously accepted
(Udalski \etal 1998a). The distance determination to the Magellanic
Clouds is one of the most important problem of the modern astrophysics
as the extragalactic distance scale is tied to the LMC distance
(Kennicutt, Freedman and Mould 1995).

The red clump method of distance determination is a single-step method
which employs the mean {\it I}-band magnitude of the red clump stars as
a standard candle. The red clump is formed from intermediate age
($2-10$~Gyr) He-core, H-shell burning stars in quiet and relatively long
phase of evolution. This is an equivalent of horizontal branch in old
($> 10$~Gyr) population of stars.

The main advantage of the red clump method is large number of the red
clump stars which makes their mean magnitude determination
straightforward and precise. Such stars are also very numerous in the
solar neighborhood which makes it possible to calibrate this standard
candle very precisely. Indeed, the best calibration is based on
Hipparcos measurements of a few hundred of nearby stars ($d< 70$~pc)
which have good quality parallaxes, photometry and are practically
unaffected by extinction (Stanek and Garnavich 1998). No other standard
candle candidate can be calibrated so accurately with direct
measurements.

The main sources of error in distance determination with the red clump
method are extinction uncertainties and possible population effects
which may affect the mean {\it I}-band brightness of the red clump in
different environments. Both problems are, however,  common to all
standard candle candidates. 

In the first approximation the same mean absolute magnitude of the red
clump stars, that determined from the local Hipparcos sample, was used
for distance determination to the Magellanic Clouds and other objects
(Paczy{\'n}ski and Stanek 1998, Stanek and Garnavich 1998, Udalski \etal
1998a, Stanek, Zaritsky and Harris, 1998). This approach was, however,
questioned by Cole (1998) and Girardi \etal (1998) who suggested large,
reaching $\approx 0.6$~mag, variations of the mean magnitude of the red
clump in objects of different age and/or metallicity based on
theoretical models.

The theoretical approach of verification of population effects on the
mean red clump luminosity suffers, however, from many assumptions which
are difficult to verify, namely helium content, mass loss, star
formation history etc. Therefore we undertook a series of observing
tests with the main goal of empirical determination how population 
effects -- metallicity and age -- can affect the mean {\it I}-band
magnitude of the red clump stars.

Udalski (1998) presented observations of the red clump stars in objects
with different metallicity which contain also another standard candle --
RR~Lyr stars. Comparison of brightness of both standard candles
indicates that  dependence of the mean {\it I}-band magnitude of the
red clump on metallicity is weak, in general about half of that of {\it
V}-band brightness-metallicity relation for RR~Lyr stars. With the most
likely slope of the RR~Lyr relation equal to 0.18 mag/dex (\cf Skillen
\etal 1993) the red clump calibration was found to be $M^{\rm
RC}_I=(0.09\pm0.03)\times{\rm [Fe/H]^{\rm RC}}-0.23\pm0.03$.

In this paper we present another test -- relation between the mean {\it
I}-band brightness of the red clump and its age. To test this effect we
observed several star clusters of different age in both LMC and SMC.
Star clusters provide an excellent opportunity for performing such a
test. First, they contain stars of the same, well defined age. Second,
by selecting clusters located in the halo of each galaxy where 
interstellar extinction is small, possible error from this source is
also small -- much  smaller than in regions closer to the centers of the
Magellanic Clouds. Finally, intermediate age clusters in both Magellanic
Clouds are known to have similar metallicity (Bica \etal 1998, Da Costa
and Hatzidimitriou 1998) thus minimizing possible influence of
metallicity on brightness. Moreover, the Magellanic Clouds provide a very
good place for such a test by assuring that the usually very uncertain
parameter -- the distance to the cluster -- is in the first
approximation the same for all objects.

\Section{Observations}

Observations of the Magellanic Cloud clusters were carried out as a
sub-project of the second phase of the OGLE microlensing survey
(Udalski, Kubiak and Szyma{\'n}ski 1997) with the 1.3-m Warsaw telescope
at the Las Campanas Observatory which is operated by the Carnegie
Institution of Washington. Single chip "first generation" CCD camera
with $2048\times 2048$ pixel SITe thin chip was used. All frames were
obtained in the normal "still frame" mode with the "medium" reading
speed. The pixel size of the CCD detector was 0.417 arcsec/pixel -- full
frame covered about $14.2\times 14.2$ arcmins on the sky.  More details
about the instrumental setup can be found in Udalski, Kubiak and
Szyma{\'n}ski (1997).

\MakeTable{lccccccc}{12.5cm}{Observed Magellanic Cloud star clusters}
{
\hline
\noalign{\vskip2pt}
\multicolumn{8}{c}{Large Magellanic Cloud}\\
\noalign{\vskip2pt}
\hline
\noalign{\vskip2pt}
&&&\multicolumn{2}{c}{Exp. time}&\multicolumn{2}{c}{No. of}&$R_{\rm CL}$\\
Name& RA(J2000)& DEC(J2000)&\multicolumn{2}{c}{(sec)}&\multicolumn{2}{c}{frames}&[\arcs]\\
&&&$V$&$I$&$V$&$I$&\\
\noalign{\vskip2pt}
\hline
\noalign{\vskip2pt}
SL388      & 5\uph20\upm06\ups & $-63\arcd28\arcm50\arcs$ & 300 & 300 & 7 & 8 & 42\\
SL663      & 5\uph42\upm30\ups & $-65\arcd21\arcm49\arcs$ & 300 & 300 & 5 & 5 & 52\\
NGC~2121   & 5\uph48\upm13\ups & $-71\arcd28\arcm46\arcs$ & 300 & 300 & 5 & 5 & 66\\
NGC~2155   & 5\uph58\upm33\ups & $-65\arcd28\arcm45\arcs$ & 300 & 300 & 5 & 5 & 50\\
ESO121SC03 & 6\uph02\upm03\ups & $-60\arcd31\arcm25\arcs$ & 300 & 300 & 6 & 8 & 58\\
SL862      & 6\uph13\upm27\ups & $-70\arcd41\arcm45\arcs$ & 300 & 300 & 6 & 6 & 50\\
\noalign{\vskip2pt}
\hline
\noalign{\vskip2pt}
\multicolumn{8}{c}{Small Magellanic Cloud}\\
\noalign{\vskip2pt}
\hline
\noalign{\vskip2pt}
&&&\multicolumn{2}{c}{Exp. time}&\multicolumn{2}{c}{No. of}&$R_{\rm CL}$\\
Name& RA(J2000)& DEC(J2000)&\multicolumn{2}{c}{(sec)}&\multicolumn{2}{c}{frames}&[\arcs]\\
&&&$V$&$I$&$V$&$I$&\\
\noalign{\vskip2pt}
\hline
\noalign{\vskip2pt}
L1         & 0\uph04\upm00\ups & $-73\arcd28\arcm00\arcs$ & 420 & 420 & 4 & 4 & 71\\
KRON3      & 0\uph24\upm46\ups & $-72\arcd47\arcm37\arcs$ & 420 & 420 & 4 & 4 & 46\\
NGC~121    & 0\uph26\upm50\ups & $-71\arcd31\arcm00\arcs$ & 420 & 420 & 4 & 4 & 66\\
L11        & 0\uph27\upm45\ups & $-72\arcd46\arcm56\arcs$ & 420 & 420 & 4 & 4 & 38\\
NGC~339    & 0\uph57\upm48\ups & $-74\arcd29\arcm00\arcs$ & 420 & 420 & 4 & 4 & 46\\
NGC~411    & 1\uph07\upm54\ups & $-71\arcd46\arcm00\arcs$ & 420 & 420 & 4 & 4 & 33\\
L113       & 1\uph49\upm30\ups & $-73\arcd43\arcm00\arcs$ & 420 & 420 & 4 & 4 & 42\\
\hline
}

Table~1 lists clusters observed in the LMC and SMC. Observations of the
LMC clusters started on April 16, 1998 and lasted until May 6, 1998.
About $5-8$ frames in the {\it V} and {\it I}-bands were collected for
each cluster. Exposure time was 300 sec for both {\it V} and {\it
I}-bands. Observations of the SMC clusters were obtained from May 14,
1998 through May 18, 1998. Four 420 sec pairs of {\it VI}-band images
were obtained for each cluster. Observations were conducted during good
weather conditions with the median seeing of about 1.3 arcsec.

Additionally, several standard stars from the Landolt (1992) fields were
observed on seven photometric nights for transformation of the
instrumental magnitudes to the standard system.

\Section{Reductions}

All frames were de-biased and flat-fielded with the standard OGLE data
pipeline. Next, photometry of stars was derived using OGLE version of
the {\sc DoPhot} photometry program (Schechter, Saha and Mateo 1993). As
the star clusters  cover small part of the frame, $512\times 512$ pixel
subframes centered on the cluster were cut from the full frames and
reduced. Photometry of images obtained at the best seeing conditions was
used to define the instrumental system of a given band and given cluster
to which photometry of the remaining images was tied.

Instrumental photometry was calibrated in similar way as regular OGLE
data (Udalski \etal 1998b). First the aperture correction was determined
from several brightest stars on the frame. Then the "total correction"
to the instrumental photometry was derived based on observations of
standard stars and transformation to the standard system as in Udalski
\etal (1998b). "Total corrections" from a few nights were averaged and
added to the instrumental photometry when databases of measurements were
created. Finally, {\it VI}-photometry of each stellar object was
determined by averaging individual measurements and including color
terms resulting from transformation. The error of zero points of
photometry should not exceed 0.02~mag.

\Section{Red Clump in MC Clusters}

The main goal of this study is to investigate dependence of the mean
{\it I}-band  brightness of the red clump stars on age. Therefore the
sample of observed clusters included intermediate age ($2-10$ Gyr)
clusters for which age estimates have been published in the literature.
In the case of the LMC only a few clusters fall into this range of age
-- there is a well known "age gap" between many young ($<2$ Gyr)
clusters and old ($>12$ Gyr) cluster population (\cf Olszewski, Suntzeff
and Mateo 1996). Only one object with age of  9~Gyr has been found,
ESO121SC03, and it was included in the list of our targets. Recently,
Sarajedini (1998) suggested that three additional objects, namely
NGC~2121, NGC~2155 and SL663, can fall in the middle of the "age gap".
Therefore we added all three clusters to our object list. On the "young"
end of our sample of LMC clusters we selected two objects observed by
Bica \etal (1998)  for which age and metallicities were determined. As
indicated by the Bica \etal (1998) observations both clusters contain red
clumps.

\MakeTable{lrccccc}{12.5cm}{Main properties of Magellanic Cloud clusters}
{
\hline
\noalign{\vskip2pt}
\multicolumn{7}{c}{Large Magellanic Cloud}\\
\noalign{\vskip2pt}
\hline
\noalign{\vskip2pt}
Cluster    & Age &  Me    & $E(B-V)$ & $E(V-I)$ & $A_I$ & References\\
           & Gyr & [Fe/H] &        &      &     &\\
\noalign{\vskip2pt}
\hline
\noalign{\vskip2pt}
ESO121SC03 & 9.0 & $-1.0$ & 0.044  & 0.06 & 0.09 & (1)\\
SL663      & 4.7 &  --    & 0.065  & 0.08 & 0.13 & (2)\\
NGC~2155    & 4.5 & $-0.6$ & 0.052  & 0.07 & 0.10 & (2)\\
NGC~2121    & 4.5 & $-0.6$ & 0.140  & 0.18 & 0.27 & (2)\\
SL388      & 2.2 & $-0.7$ & 0.042  & 0.05 & 0.08 & (1)\\
SL862      & 1.8 & $-0.9$ & 0.119  & 0.15 & 0.23 & (1)\\
\hline
\noalign{\vskip2pt}
\multicolumn{7}{c}{Small Magellanic Cloud}\\
\noalign{\vskip2pt}
\hline
\noalign{\vskip2pt}
Cluster    & Age &  Me    & $E(B-V)$ & $E(V-I)$ & $A_I$ & References\\
           & Gyr & [Fe/H] &        &      &     &\\
\noalign{\vskip2pt}
\hline
\noalign{\vskip2pt}
NGC~121     &12.0 & $-1.5$ & 0.033  & 0.04 & 0.06 & (3)\\
L1         & 9.0 & $-1.1$ & 0.032  & 0.04 & 0.06 & (3,4)\\
KRON3      & 7.5 & $-1.1$ & 0.031  & 0.04 & 0.06 & (3,7)\\
NGC~416     & 6.6 & $-1.4$ & 0.08~~ & 0.10 & 0.16 & (4)\\
L113       & 5.3 & $-1.3$ & 0.048  & 0.06 & 0.09 & (3,4)\\
NGC~339     & 4.0 & $-1.4$ & 0.050 & 0.06 & 0.10 & (3)\\
L11        & 3.5 & $-0.7$ & 0.034  & 0.04 & 0.07 & (3)\\
NGC~419     & 3.3 &  --    & 0.08~~ & 0.10 & 0.16 & (6)\\
NGC~411     & 1.5 & $-0.9$ & 0.08~~ & 0.10 & 0.16 & (5)\\
\noalign{\vskip2pt}
\hline
\noalign{\vskip2pt}
\multicolumn{6}{p{8cm}}{(1)~Bica \etal (1998),
(2)~Sarajedini (1998),
(3)~Da Costa and Hatzidimitriou (1998),
(4)~Mighell, Sarajedini and French (1998),
(5)~Da Costa and Mould (1986),
(6)~Bica \etal (1986),
(7)~Rich, Da Costa and Mould (1984).}\\
}

In the case of the SMC  the sample of clusters is larger and covers the
intermediate age range more uniformly. We also added to our list the
oldest cluster in the SMC -- NGC~121 which is about 12~Gyr old and it is
the only one in the SMC which possesses a few RR~Lyr stars. Two
additional objects with determined age, NGC~416 and NGC~419, were
observed during the regular OGLE observations and were already presented
in the Catalog of Star Clusters in the SMC (Pietrzy{\'n}ski \etal 1998).
We included both objects in this study.

The list of star clusters from the Magellanic Clouds analyzed in this
paper is given in Table~2. Basic parameters, age and metallicity of each
object are also included there. Age is the average from values found in
the literature. We adopted the age scale in which ESO121SC03 in the LMC
and L1 in the SMC are 9 Gyr old. Also interstellar reddening $E(B-V)$ to
each cluster is given. The reddening was determined from the COBE/IRAS
maps of Schlegel, Finkbeiner and Davis (1998) with exception of NGC~416,
NGC~419 and NGC~411 in the SMC which are located close to the central
parts of the SMC where COBE/IRAS maps are not reliable. For these
objects reddening $E(B-V)=0.08$ was assumed based on determination of
Mighell, Sarajedini and French (1998) for NGC~416. Majority of clusters
from our sample is located in the halo of both galaxies where extinction
is small and therefore potential errors from this source are also small.
Only NGC~2121 in the LMC is located dangerously close to the region
where COBE/IRAS maps are not reliable. However, its value of reddening
from the COBE/IRAS maps seems to be reasonable, and we adopted it though
it might be slightly overestimated. In further analysis we assumed
standard extinction law with $A_I$ extinction equal to $1.96\times
E(B-V)$ and $E(V-I)$ reddening: $1.28\times E(B-V)$ (Schlegel,
Finkbeiner and Davis 1998).

Figs.~1 and 2 present color-magnitude diagrams (CMDs) of clusters
from our sample. Only stars located closer than $R_{\rm CL}$ from the
center of the cluster were included. Adopted values of $R_{\rm CL}$ are
listed in Table~1. Additionally observations were filtered with
$\sigma_{\rm max}$ filter passing stars with standard deviation smaller
than maximum standard deviation for a given {\it I}-band magnitude and
thus eliminating stars with poor photometry and variable objects.
Maximum standard deviation \vs magnitude limit of non-variable
objects was determined from photometry of all stars. No statistical
cleaning of the CMDs was performed because the $R_{\rm CL}$ was selected
to include mostly the central part of each cluster and majority of
clusters is located in the empty halo regions where contamination of
CMDs by field stars should be small.

{\it I} \vs $(V-I)$ CMDs of the LMC clusters (Fig.~1) are presented
according to decreasing age order. Although our CMDs are not deep enough
and we do not  attempt to determine the  age by isochrone fitting, it is
evident that the ESO121SC03 cluster is the oldest from the sample while
SL862 the youngest one. It is not obvious that three objects proposed by
Sarajedini (1998) to be in the middle of the "age gap", namely  SL663,
NGC~2155 and NGC~2121, are indeed about $4-5$ Gyr old but we will assume
that age as based on much deeper HST observations.

Fig.~2 presents CMDs of the SMC star clusters. Again the clusters
are shown in decreasing age order. Morphology of the CMDs suggests that
the assumed  age sequence of our sample is correct with NGC~121 being
the oldest and NGC~411 the youngest object. The age of the remaining
clusters seems to be in agreement with observed position of the red
giant branch and/or the main sequence turn-off region. The red clump is
well visible although in some less populous objects (SL388 in the LMC,
L11 and L113 in the SMC) the number of red clump stars is small.

The mean {\it I}-band magnitude of the red clump stars was determined in
similar way as in earlier analysis (\eg Udalski \etal 1998a). First,
histograms of the {\it I}-band luminosity of stars in 0.07~mag bins were
constructed. Stars from the following range were included in these
histograms: $17.0<I<19.0$ and $0.8<(V-I)<1.2$ for the LMC and
$17.5<I<19.5$ and $0.7<(V-I)<1.1$ for the SMC. Then a Gaussian
representing the red clump stars distribution superimposed on the second
order polynomial function approximating the stellar background was
fitted:

$$n(I)=a+b(I-I^{\rm max})+c(I-I^{\rm max})^2+
\frac{N_{RC}}{\sigma_{\rm RC}\sqrt{2\pi}} \exp\left[-\frac{(I-I^{\rm
max})^2}{2\sigma^2_{\rm RC}}\right]\eqno(1)$$

Figs.~3 and 4 show histograms with the fitted function given by Eq.~(1)
for the LMC and SMC clusters, respectively. One can note that in spite
of small number of stars in some cases, the red clump is well pronounced
in each cluster and the mean {\it I}-band brightness of the red clump
can be determined with good accuracy. The number of stars used for
histograms, $N_*$, mean red clump brightness, $I^{\rm max}$, standard
deviation of the fitted Gaussian, $\sigma_{\rm RC}$, statistical error,
$\sigma^{\rm STAT}$, and systematic error from extinction uncertainty,
$\sigma^{\rm SYS}$, are listed in Tables~3 and 4. Figs.~5 and 6 present
enlargements of CMDs around the red clump of each cluster with dotted line
showing its mean magnitude.

\MakeTable{lrcccc}{12.5cm}{Red clump in LMC clusters}
{
\hline
\noalign{\vskip2pt}
Cluster    &$N_*$&$I^{\rm max}$ & $\sigma_{\rm RC}$ & $\sigma^{\rm STAT}$ &
$\sigma^{\rm SYS}$ \\
\noalign{\vskip2pt}
\hline
\noalign{\vskip2pt}
ESO121SC03 &  36 & 18.03 & 0.05 & 0.01 & 0.03 \\
SL663      &  63 & 18.02 & 0.09 & 0.02 & 0.03 \\
NGC~2155    &  82 & 17.92 & 0.06 & 0.02 & 0.03 \\
NGC~2121    & 221 & 18.10 & 0.09 & 0.01 & 0.04 \\
SL388      &  46 & 17.96 & 0.10 & 0.03 & 0.03 \\
SL862      &  49 & 18.16 & 0.05 & 0.01 & 0.04 \\
\noalign{\vskip2pt}
\hline
}

Extinction-free, mean {\it I}-band magnitude of the red clump stars,
$I_0$, and their mean color $\langle(V-I)_0\rangle$ for the LMC and SMC
clusters are listed in Table~5. They were determined using extinction
and reddening values listed in Table~2. In the case of the LMC which is
believed to be located almost "face-on" no additional correction to
$I_0$ was applied. In the case of the SMC the situation is different and
it is believed that this galaxy is tilted to the line-of-sight and has
non-negligible depth (\cf Caldwell and Laney 1991). As some of the
observed clusters are located a few degrees from the SMC center
the geometric correction cannot be neglected.

However, OGLE observations of the SMC seem to suggest that the depth
effect might be smaller than previously assumed. The mean brightness of
the red clump and RR Lyr stars from two fields located at opposite --
East and West -- ends of the bar show no brightness difference at the
level of 0.02~mag (Udalski \etal 1998, Udalski 1998). Observations in
many other lines-of-sight in the SMC halo are currently collected during
the regular OGLE observations. They should reveal precise geometry of
the SMC when sufficient number of images is obtained.

There are two lines-of-sight in the halo of the SMC to which we already
can determine geometric correction. Fortunately both are close to some
of our clusters and enable us to determine geometric correction with
good precision based on observations obtained with modern techniques.
First region is located in the direction towards NGC~121. This cluster is
known to contain four RR~Lyr stars. All four objects were observed with
the CCD technique by Walker and Mack (1988)  who determined the mean
magnitude of the entire sample to be $V=19.59$. The region around this
cluster was also searched for RR~Lyr stars with photographic survey by
Graham (1975). The mean {\it V}-band magnitude of about 70  RR~Lyr stars
from that area was found to be $V=19.57$ (Graham 1975). Although the
calibration of photographic plates is often not very accurate we could
verify both values with our CCD observations of the NGC~121 cluster.
Therefore we repeated reductions of NGC~121 and obtained photometry
of the entire $14.2\times 14.2$ arcmin field around the cluster. We
identified four RRab~Lyr stars belonging to the cluster and six
additional objects from the field. Then we calculated the mean  {\it
V}-band magnitudes of both samples. Although the number of individual
observations is small -- only four images -- our previous experience
shows that the mean value of magnitude of our sample containing a few
RR~Lyr stars should be reasonable (Udalski 1998).

The mean magnitude of four RR~Lyr stars in NGC~121 was found to be
$V=19.59\pm0.03$~mag and that of six field RR~Lyr stars practically
identical: $V=19.60\pm0.03$. This is in excellent agreement with Walker
and Mack's and Graham's values. Correcting for extinction in that
direction we obtain $V_0=19.49\pm0.04$ which compared to the mean
magnitude of the RR~Lyr stars in the bar: $V_0^{BAR}=19.41$ (Udalski
1998) suggests that the region of SMC around NGC~121 is about 0.08~mag
behind the central parts of the galaxy.

The second region where geometric correction can be determined lies
towards the galactic globular cluster 47 Tuc (${\rm
RA(J2000)}=0\uph24\upm, {\rm DEC(J2000)}=-72\arcd05\arcm$). Ten RRab Lyr
stars belonging to the SMC were found in that direction during the first
phase of the OGLE project (Kaluzny \etal 1998a). Additional six objects
were found by Kaluzny \etal (1997). The latter sample contains three
stars in common with the OGLE sample but the light curves have much
worse phase coverage. We determined the mean intensity weighted
magnitudes of each object from both samples and then determined the mean
{\it V}-band magnitudes of each sample. Results were almost identical:
$V=19.68$ mag and $V=19.69$ mag for the OGLE and Kaluzny \etal (1997)
samples, respectively. Correcting for extinction, $E(B-V)=0.04$ (Kaluzny
\etal 1998b), we obtain $V_0=19.55\pm0.03$ which leads to conclusion
that this region of the SMC is located about 0.14~mag behind the central
parts.

\MakeTable{lrccccrc}{12.5cm}{Red clump in SMC clusters}
{\hline
\noalign{\vskip2pt}
Cluster & $N_*$ & $I^{\rm max}$ & $\sigma_{\rm RC}$ & $\sigma^{\rm STAT}$ & 
$\sigma^{\rm SYS}$ & Geo Cor &  $\sigma^{\rm GeoCor}$ \\
\noalign{\vskip2pt}
\hline
\noalign{\vskip2pt}
NGC~121 & 214 & 18.86 & 0.09 & 0.01 & 0.03 & $-0.08$ & 0.04 \\
L1     & 163 & 18.44 & 0.05 & 0.01 & 0.03 & $-0.20$ & 0.06 \\
KRON3  & 203 & 18.46 & 0.08 & 0.01 & 0.03 & $-0.13$ & 0.04 \\
NGC~416 &  73 & 18.61 & 0.11 & 0.03 & 0.05 & $ 0.00$ & 0.00 \\
L113   &  53 & 18.22 & 0.06 & 0.02 & 0.03 & $ 0.20$ & 0.06 \\
NGC~339 & 104 & 18.43 & 0.15 & 0.02 & 0.03 & $ 0.00$ & 0.00 \\
L11    &  49 & 18.50 & 0.09 & 0.02 & 0.03 & $-0.13$ & 0.04 \\
NGC~419 & 235 & 18.43 & 0.16 & 0.02 & 0.05 & $ 0.00$ & 0.00 \\
NGC~411 &  81 & 18.47 & 0.16 & 0.02 & 0.05 & $ 0.00$ & 0.00 \\
\noalign{\vskip2pt}
\hline
}

With this information we might correct observed mean luminosity of the
red clump in SMC clusters for geometric effects. For clusters located in
or close to the bar: NGC~416, NGC~419, NGC~411 and NGC~339 correction is
negligible. Correction for NGC~121 was determined above and it is equal
to $-0.08\pm0.04$~mag. KRON3 and L11 clusters are located near 47~Tuc
region -- somewhat towards the bar. Thus we assumed correction of
$-0.13\pm0.04$ mag for both clusters. L1 is located considerably farther
to the West. Therefore extrapolated correction of $-0.20\pm0.06$~mag was
adopted for this cluster. For L113, located on the opposite eastern side
of the SMC in more or less similar angular distance from the SMC bar as
L1, we assumed correction of $0.20\pm0.06$~mag based on Gardiner and
Hatzidimitriou (1992) estimate. It should be noted that geometric
corrections for both most uncertain locations of L1 and L113 clusters
will be determined more precisely when new OGLE data being gathered in
the neighboring regions are completed.

\MakeTable{lccc}{12.5cm}{Extinction-free photometry of the red clump in MC clusters}
{
\hline
\noalign{\vskip2pt}
\multicolumn{4}{c}{Large Magellanic Cloud}\\
\noalign{\vskip2pt}
\hline
\noalign{\vskip2pt}
Cluster   & $I_0$ & $\sigma^{\rm TOT}$ & $<(V-I)_0>$\\
\noalign{\vskip2pt}
\hline
\noalign{\vskip2pt}
ESO121SC03 & 17.94 & 0.04 & 0.89\\
SL663      & 17.89 & 0.04 & 0.92\\
NGC~2155   & 17.82 & 0.04 & 0.89\\
NGC~2121   & 17.83 & 0.05 & 0.86\\
SL388      & 17.88 & 0.05 & 0.93\\
SL862      & 17.93 & 0.05 & 0.87\\
\hline
\noalign{\vskip2pt}
\multicolumn{4}{c}{Small Magellanic Cloud}\\
\noalign{\vskip2pt}
\hline
\noalign{\vskip2pt}
Cluster   & $I_0$ & $\sigma^{\rm TOT}$ & $<(V-I)_0>$\\
\noalign{\vskip2pt}
\hline
\noalign{\vskip2pt}
NGC~121 &  18.72 & 0.06 & 0.76\\
L1      &  18.18 & 0.08 & 0.82\\
KRON3   &  18.27 & 0.06 & 0.85\\
NGC~416 &  18.45 & 0.06 & 0.81\\
L113    &  18.33 & 0.08 & 0.86\\
NGC~339 &  18.33 & 0.04 & 0.85\\
L11     &  18.30 & 0.06 & 0.87\\
NGC~419 &  18.27 & 0.05 & 0.82\\
NGC~411 &  18.31 & 0.05 & 0.81\\
\noalign{\vskip2pt}
\hline
}

Geometric corrections for the SMC clusters with errors are listed in
Table~4. Final mean {\it I}-band magnitude of the red clump stars in the
SMC clusters, $I_0$, is listed in Table~5. The main component of the
error budget, $\sigma^{\rm TOT}$, comes from uncertainty of geometrical
correction and extinction uncertainty for the SMC clusters and
extinction uncertainty for the LMC clusters.

\Section{Discussion}

Fig.~7 presents dependence of the mean {\it I}-band brightness of the
red clump stars on cluster age for the LMC and SMC. Filled dots
represent clusters of intermediate age -- 1.5--10~Gyr. The most evident
conclusion from Fig.~7 is that the mean {\it I}-band brightness of the
red clump for intermediate age clusters is constant. For the LMC
clusters the mean red clump brightness is $I_0=17.88$~mag with the
standard deviation of only 0.05~mag. In the case of the SMC clusters
$I_0=18.31$~mag and the standard deviation is somewhat larger: 0.07~mag.
However, the main contributors are NGC~416, which red clump is
relatively poorly defined, and L1, to which the geometrical correction
is the most uncertain. Nevertheless, even in these extreme cases the
mean magnitude of the red clump of both clusters does not deviate by
more than $2\sigma$ from the mean value. In general some deviations from
the mean value could be expected for two reasons: slight differences in
metallicity between clusters and different location within each galaxy.
Both effects seem to be, however, small, not exceeding a few hundredths
of magnitude.

The oldest cluster in the SMC -- NGC~121 requires special attention. Its
red clump, which is shown in Fig.~7 as an asterisk, is considerably
fainter (by about 0.4~mag), than those of the remaining clusters. There
are two possible explanations. First, the cluster is significantly
farther away than the main part of the SMC in this direction. The second
possibility is that the red clump of NGC~121 is intrinsically fainter
than those of the remaining younger clusters.  Fig.~8 presents the CMD
of the entire NGC~121 field. A star indicates the mean position of four
RR~Lyr stars belonging to this cluster.

As can be seen two red clumps are clearly distinguishable in Fig.~8 --
one of the field stars and the fainter one of NGC~121. The mean {\it
I}-band magnitude of the red clump of field stars is $I=18.46$ which
leads to the extinction-free magnitude: $I_0^{\rm NGC121FIELD}=18.40$.
Comparing to the mean magnitude of the red clump from the bar regions
$I_0^{\rm BAR}=18.33$ (Udalski \etal 1998a) one can conclude that this
region is 0.07~mag behind the SMC center, in very good agreement with
the geometric correction determined in Section~4 from RR~Lyr stars.

There is a striking resemblance of the CMD of the NGC~121 field with the
CMD of the Carina dwarf galaxy presented in Fig.~5 by Udalski (1998).
The Carina dwarf galaxy is known to contain an old (about 15~Gyr)
population of stars and also much younger -- intermediate age stars, 7
and 3~Gyr old. The older population of He-burning stars forms a well
developed horizontal branch (HB) while intermediate age stars -- well
pronounced red clump. The red clump is approximately 0.3--0.4~mag
brighter than the red part of the HB (defined as the part between the
red giant branch and location of the RR~Lyr stars). The situation is
analogous in NGC~121 field. This cluster is significantly older than the
remaining ones in the SMC and its age is comparable, though somewhat
younger, to that of the oldest galactic or LMC clusters. Its red clump
should be then  interpreted as the red part of the HB which starts to
develop in older clusters. Indeed, Fig.~8 shows that it is extended
towards the RR~Lyr objects present in this cluster. Small sample of
RR~Lyr stars in NGC~121 confirms that the object is one of the youngest
in the group of the old clusters. The mean luminosity of the red
clump/red HB of NGC~121 is fainter by about 0.4~mag than the red clump
of the intermediate age population of field stars, fully analogous to
the Carina dwarf galaxy.

To look for similar examples of brightness difference between the old
and intermediate age population of He-burning stars we browsed through
literature looking for old clusters in the LMC with {\it VI} photometry.
Unfortunately only NGC~2210 has both reasonably populated red HB and
reliable photometry (Fig.~8, Reid and Freedman 1994). Crude estimate of
the mean {\it I}-band magnitude of the red HB of this cluster is
$I=18.25\pm0.1$~mag \ie $I_0=18.1\pm0.1$. The asterisk in the LMC panel
of Fig.~7 indicates position of NGC~2210. Again the situation is similar
to the SMC and Carina galaxy. The red HB is fainter by about 0.25~mag 
than the red clump of intermediate age population. The difference might
be even 0.1~mag larger as the difference between metallicity of the
intermediate age and old cluster populations is about twice that big for
the LMC than SMC, making the red HB of the old clusters brighter in the
LMC.

Another argument in favor of our interpretation of the NGC~121 red clump
provides the CMD of the Fornax dwarf galaxy (Stetson, Hesser and
Smecker-Hahne 1998). This galaxy also contains variety of stellar
populations with both old and intermediate age population similar to the
Carina dwarf galaxy. The CMD of the Fornax galaxy (Fig.~5, Stetson,
Hesser and Smecker-Hahne 1998) looks basically identical as those of the
Carina galaxy and NGC~121 field -- the intermediate age red clump is
brighter than the red HB of the old population. More quantitative
conclusion cannot be unfortunately drawn because the Stetson, Hesser and
Smecker-Hahne (1998) observations were made in different bands ({\it
BR}).

Concluding, we tend to interpret the fainter red clump in the NGC~121 as
the red part of just developing HB rather than caused by different
distance to the cluster. In the latter case the cluster would have to be
located about 10~kpc ($\approx 20$\%) behind the SMC which is not likely
as no other cluster from our sample deviates that much from the mean
distance. Larger distance can also be ruled out by the mean brightness
of the RR~Lyr stars belonging to the cluster. It is basically the same
as that of the field RR~Lyr stars from this region and only a few
hundredth of magnitude fainter than that of RR~Lyr stars from central
regions of the SMC. Thus the red clump/red HB in NGC~121 is
intrinsically fainter by about 0.4~mag than in younger clusters. It
should be noted that this effect is not caused by metallicity
differences as the metallicity of NGC~121 is practically the same as
many younger SMC clusters (Table~2). Also the old and intermediate age
population in the Carina dwarf galaxy, where similar effect was
observed, have very similar metallicity (Udalski 1998).

Results of observations of the mean {\it I}-band luminosity of the red
clump in clusters of different age in different galaxies, LMC and SMC,
are very consistent and support the following empirical picture. The
red clump stars in young clusters ($<1.5-2.0$~Gyr) are brighter
when the object is younger and their brightness stabilizes gradually  with
increasing age. A sample of 11 young clusters in the LMC was observed by
Corsi \etal (1994) and the top panel of their Fig.~36 clearly shows that
behavior. Unfortunately photometry of Corsi \etal (1994) was obtained in
the {\it BV}-bands so we cannot include their results in our Fig.~7.
Also CMDs of the field stars in both Magellanic Clouds provide another 
example of such a behavior (\eg Zaritsky and Lin 1997, Beaulieu and
Sackett 1998, Udalski \etal 1998a). A well pronounced tail of brighter
stars, an extension to the red clump called sometimes vertical red clump
is clearly seen in these diagrams.

Observations of clusters of intermediate age (2--10~Gyr) presented in
this paper clearly show that the mean {\it I}-band luminosity of the red
clump is constant at the level better than 0.05~mag over the entire
range of ages. Then, for objects older than about 10~Gyr the red clump
becomes more extended starting to form the red part of the HB
characteristic for old population. Simultaneously the mean {\it I}-band 
magnitude of the red clump, becoming now the red part of the HB fades
and for objects older than 12~Gyr (\eg NGC~121) it is about
$0.3-0.4$~mag fainter than the red clump of intermediate age population.
This effect is clearly seen, for example, in M31 (Fig.~4, Stanek and
Garnavich 1998). Blue part of the red clump of field stars in this
galaxy converts smoothly into the red HB which brightness fades by about
0.2~mag. For older objects the brightness of the red HB  likely remains
more or less constant though we have only two crude quantitative
examples -- the Carina galaxy and NGC~2210. In older objects the red
part of the HB becomes poorly populated and it is bent in the {\it
I}-band towards fainter magnitudes which makes precise determination of
its mean magnitude practically impossible.

The observed behavior of the red clump mean {\it I}-band luminosity sets
a very important limitation on the red clump method of distance
determination. One must make sure that the red clump in  object to which
the distance is to be determined is formed by stars of intermediate age,
\ie $2-10$~Gyr. Otherwise the red HB taken as the intermediate age red
clump, as possible in the case of NGC~121, can lead to erroneous
distance modulus estimation by as much as 0.4~mag.

The age of the target object can be determined from a deep CMD. However,
it seems that the dereddened $(V-I)_0$ color of the red clump might
provide a crude test whether the red clump method can be applied.
Observations of the Magellanic Clouds show that it seems reasonable to
assume the mean color of the red clump  $\langle(V-I)_0\rangle > 0.8$ as
the range where the red clump method can safely be applied. Similar
constraint can also be deduced from observations of M31 and Carina
galaxies (Fig.~4, Stanek and Garnavich 1998; Fig.~5,  Udalski 1998). A
bluer red clump suggests that it is likely formed from older stars which
mean brightness is lower. Having this limitation in mind the red clump
method of distance determination can be applied to several galactic
clusters which contain red clump (\eg Sarajedini, Lee and Lee 1995). The
distance to 47~Tuc which seems to contain the red clump rather than red
HB as indicated by its mean $(V-I)_0$ red clump color has already been
determined by Kaluzny \etal (1998b): $m-M=13.32\pm0.05$ mag (which with
a small correction for metallicity, according to calibration of Udalski
1998, increases to $m-M=13.38$).

The independence of the mean {\it I}-band brightness of the red clump
stars on age for the intermediate ages makes it possible to use the red
clump as a very good brightness reference. The mean {\it I}-band
magnitude of the red clump determined from clusters in both Magellanic
Clouds is in good agreement with that derived in fields located close to
the bar where extinction is much larger: $I_0^{\rm BAR}=17.85$ \vs
$I_0=17.88$ for the LMC and $I_0^{\rm BAR}=18.33$ \vs $I_0=18.31$ for
the SMC. This independent determination of $I_0$ confirms earlier
results of Udalski \etal (1998a) and Stanek, Zaritsky and Harris (1998).
It also indicates that extinction estimates used in all these
determinations and based on the following methods: COBE/IRAS maps
(Schlegel, Finkbeiner and Davis 1998), classical OB star determinations
(Harris, Zaritsky and Thompson 1997, Oestreicher and Schmidt-Kaler 1996)
and NGC~416 cluster determination (Mighell, Sarajedini and French 1998)
are consistent.

Determination of the mean red clump brightness from the red clump of
clusters located mostly in low extinction parts of the halo of the LMC
and SMC enables us to recalculate the distance to both galaxies based on
this new independent data set. Assuming the mean metallicity of clusters
to be $-0.8$~dex and $-1.2$~dex for the LMC and SMC, respectively, the
mean brightness of the red clump: $I_0=17.88\pm0.05$ mag and
$I_0=18.31\pm0.07$ mag and brightness-metallicity relation given by
Udalski (1998) the resulting distance moduli are $m-M=18.18\pm0.06$ mag
and $m-M=18.65\pm0.08$ mag for the LMC and SMC, respectively. These
figures are in very good agreement with previous red clump
determinations based on bar field stars, $m-M=18.13\pm0.07$ mag and
$m-M=18.63\pm0.07$ (Udalski 1998), confirming the short distance scale
to both Magellanic Clouds.

Summarizing this series of papers on the red clump method of distance
determination we conclude that there is no ideal standard candle. All
proposed and used candidates are more or less population affected and
require additional calibration. However, the mean {\it I}-band
brightness of the red clump stars seems to be the least affected by
population effects and it is now well empirically tested. The red clump
brightness is independent of age for intermediate age range ($2-10$~Gyr)
as shown in this paper. Dependence on metallicity is weak, most likely
only 0.09 mag/dex or in general approximately half of {\it V}-band
brightness-metallicity dependence for RR~Lyr stars (Udalski 1998). The
red clump stars are the only standard candle candidates calibrated
directly based on a few hundred of nearby stars with high precision
parallaxes measured by Hipparcos. Therefore unless  an additional,
unknown factor affecting the mean luminosity of the red clump stars in
different objects exists, the red clump stars seem to be the most
promising standard candle candidate. The only limitation comes from
possible mimicking of the red clump by somewhat older objects and
younger stars. Thus one has to confirm that the red clump in an
investigated object consists of stars of intermediate age (2--10~Gyr).

\renewcommand{\TableFont}{\scriptsize}
\MakeTable{p{2cm}p{3.1cm}p{3.1cm}p{3.1cm}}{12.5cm}{Comparison of
standard candle candidates}
{
\hline
\noalign{\vskip3pt}
                      &  Cepheids              & RR Lyr                & Red Clump stars\\
\noalign{\vskip3pt}
\hline
\noalign{\vskip3pt}
\raggedright{Typical number of stars in object  }&   
Small.                              & 
Small.                              & 
Very large.\\
\noalign{\vskip5pt}
Absolute calibration  & 
                      &
                      &
                      \\
\noalign{\vskip5pt}
\raggedright{a) Hipparcos trigonometric parallaxes        }&
Uncertain.          &
Uncertain.          &
\parbox[t]{3.1cm}{\raggedright Very good: more than 200 stars from $d<70$~pc.} \\
\noalign{\vskip2pt}
\raggedright{b) ground based        }&
\raggedright{Uncertain. Only $20-30$ objects with distances based on galactic
clusters. Problems with the galactic distance scale. }&
\raggedright{Reasonable. Statistical parallaxes based on radial velocities and proper
motions of more than 150 galactic objects. }&
Not necessary.\\
\noalign{\vskip5pt}
\raggedright{Population Effects    }& 
\raggedright{Possible dependence of the zero point of period luminosity 
relation on metallicity. }& 
\raggedright{Calibrated $M_V$-metallicity relation. }& 
\parbox[t]{3.1cm}{\raggedright Weak, calibrated dependence $M_I$-metallicity. 
No dependence  on age.}\\
\noalign{\vskip5pt}
Limitation            &  
$P<100$~days.         & 
                      & 
\parbox[t]{3.1cm}{\raggedright Intermediate age $2-10$~Gyr stars.}\\
\noalign{\vskip5pt}
LMC calibration       & 
\raggedright{Based on only $20-50$ stars. Inhomogeneous data set.}& 
\raggedright{More than hundred stars with good photometry and 
extinction determination.}& 
\parbox[t]{3.1cm}{\raggedright Several thousand stars with good photometry and 
extinction determination.}\\
\noalign{\vskip5pt}
\raggedright{LMC distance mo\-du\-lus }& 
\raggedright{${18.50\pm0.10}$.}& 
\raggedright{${18.08\pm0.16}$.}& 
\parbox[t]{3.1cm}{\raggedright $18.18\pm0.06$.}\\
\noalign{\vskip3pt}
\hline}

\newpage

In Table~6 we compare the most important  properties of three basic
standard candle candidates, namely Cepheids, RR~Lyr stars and red clump
stars  used as the first step in the extragalactic distance scale ladder
to show superiority of the latter. We hope the distance scale based on
the red clump method will be confirmed in the future by independent
methods like the detached eclipsing binaries (Paczy{\'n}ski 1997).

Photometry of the Magellanic Cloud clusters is available from the OGLE
Internet  archive -- 
{\it ftp://sirius.astrouw.edu.pl/ogle/ogle2/clusters/misc}.

{\bf Acknowledgements.} We would like to thank Prof.\ Bohdan Paczy\'nski
for suggestion of using MC clusters for testing the red clump stars
properties, many encouraging and stimulating discussions and help at all
stages of the OGLE project. We thank Dr.\ J.~Kaluzny for providing us
with light curves of RR~Lyr stars from 47 Tuc field and Drs.\
M.~Szyma{\'n}ski and K.Z.~Stanek for many important remarks and
comments.  Part of observations  analyzed in this paper was carried out
by Dr.\ M.~Szyma{\'n}ski. The  paper was partly supported by the Polish
KBN grant 2P03D00814 to A.\ Udalski.  Partial support for the OGLE
project was provided with the NSF grant AST-9530478 to B.~Paczy\'nski.

\newpage

\centerline{\bf Figure captions}

\vspace{1cm}

\noindent
Fig.~1. Color-magnitude diagrams of the LMC star clusters.

\noindent
Fig.~2. Color-magnitude diagrams of the SMC star clusters.

\noindent
Fig.~3. Luminosity function of the red clump stars in the LMC clusters.
Bins are 0.07~mag wide. The solid line represents the best fit of a
Gaussian superimposed on the parabola function (Eq.~1).

\noindent
Fig.~4. Luminosity function of the red clump stars in the SMC clusters.
Bins are 0.07~mag wide. The solid line represents the best fit of a
Gaussian superimposed on the parabola function (Eq.~1).

\noindent
Fig.~5. Enlargement around the red clump of the color-magnitude diagram
of the LMC clusters. Mean brightness of the red clump is shown by dotted
line.   

\noindent
Fig.~6. Enlargement around the red clump of the color-magnitude diagram
of the SMC clusters. Mean brightness of the red clump is shown by dotted
line.

\noindent
Fig.~7. Mean {\it I}-band brightness of the red clump as a function of
age for the LMC (upper panel) and SMC (lower panel). Filled dots
represent clusters of intermediate age ($2-10$~Gyr) while asterisks
older clusters.

\noindent
Fig.~8. Color-magnitude diagram of NGC~121 and $14.2\times 14.2$ arcmin
field around the cluster. The star indicates the mean location of the
four RR~Lyr stars found in NGC~121.

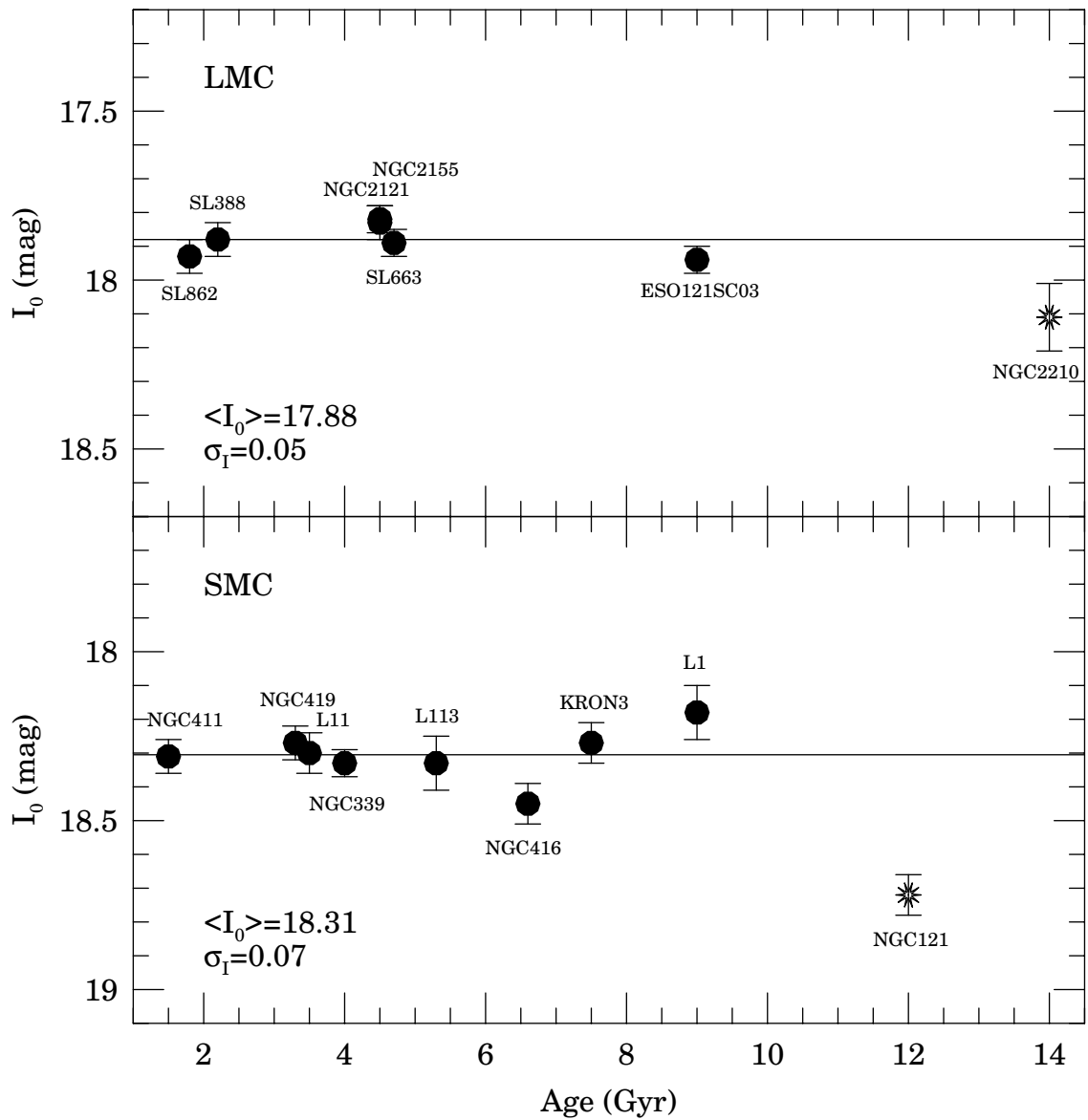
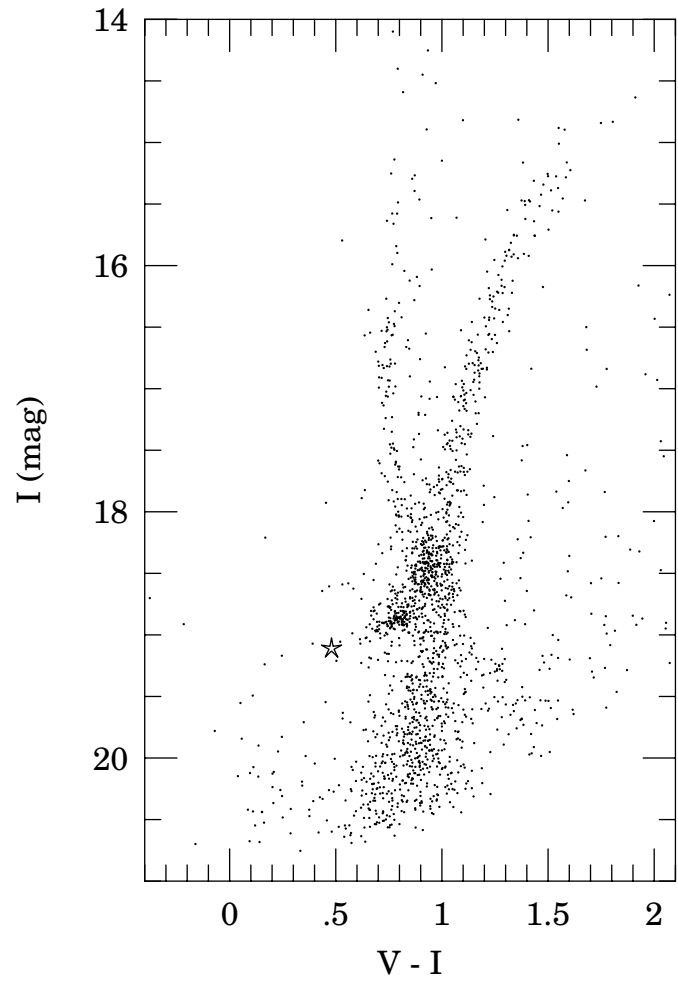
\end{document}